\documentclass[12pt]{amsart}
\usepackage{amsmath,amsthm,epsfig,amssymb}
\usepackage{amsfonts,amscd}
\usepackage[all]{xy}
\usepackage{young}

\theoremstyle{definition}
\newtheorem{Def}[subsection]{Definition}
\newtheorem{Axiom}[subsection]{Axiom}

\newtheorem{proposition-definition}[subsection]{Proposition-Definition}

\begin{normalsize} \end{normalsize}

\newcommand{\RR}{{\mathbb R}}
\newcommand{\ZZ}{{\mathbb Z}}

\newcommand{\NN}{{\mathbb N}}

\newcommand\tr{{\mathit Tr}}
\numberwithin{equation}{section}

\author{Werner Nahm}

\subjclass{01A99, 81T99}

\begin{document}

\title{A Touch of Quantum Field Theory}
 
\maketitle

\section{Introduction} \setcounter{page}{1}

In Srinivasa Ramanujan’s work there are at least three overlaps with quantum field theory, namely his treatment of divergent sums (chapter 6 of his second notebook \cite{Berndt1}), the Rogers-Ramanujan functions (chapter 16 \cite{Berndt2}), and mock theta series (Lost Notebook \cite{AndrewsBerndt}). But he died five years before even quantum mechanics was discovered. How was this possible? Did he have precocious insights that even might help with the formulation of a true theory of quantum gravity? Or are we faced with a superficial phenomenon with some simple explanation? For the moment, the question remains open, but it is certain that Ramanujan’s legacy may help mathematicians to put quantum field theory (QFT) at its proper place in mathematics. It is a framework like group theory. Special systems that fit in this framework are called QFTs, or models, for rethoric variation. Nobody would call a group a group theory, but physics jargon has its quaint sides. 

A full explanation of the mock theta story requires an extensive discussion of supersymmetry. This would need more time and space than available. On the other hand, the Rogers-Ramanujan functions are a central object in the simplest non-trivial QFT. In the following pages this statement will be made more precise. The introduction provides some terminology and context. The second section presents prototypical QFTs. It explains their connection with divergent sums and uses their proven or conjectured properties for definitions and tentative conclusions. The arguments need improvement, but they are sufficiently simple and transparent to allow any mathematician to work with them, or on them. The third section presents the Rogers-Ramanujan QFT. The fourth section gives some insight into the role of mock modular forms in certain QFTs and ends with speculations concerning the questions asked above. No physics background is assumed, though many side remarks are addressed at readers who have it. Perhaps some mathematicians will be motivated to explore QFT in their own way and perhaps they will find more links with Ramanujan’s ideas.

QFT comes in two versions, euclidean and Minkowskian. The latter is fundamental for particle physics, the former explains many phenomena in the physics of condensed matter. Generic Minkowskian theories have euclidean partners, but euclidean QFTs have to satisfy a positivity condition to admit a Minkowskian version. A commonly used word for this condition is unitary. The Minkowskian partners can be obtained by continuing a space coordinate to imaginary values, so that it becomes time. When differential equations appear, they are elliptic in the euclidean and hyperbolic in the Minkowskian version.  Minkowskian QFT relates to operator algebras and distributions, which appear to be far from Ramanujan’s train of thought. Moreover, the Rogers-Ramanujan QFT is not unitary. Thus we will focus attention on the euclidean case. 

Fifty years ago, Dyson regretted that “the marriage between mathematics and physics, which was so enormously fruitful in past centuries, has recently ended in divorce” (\cite{Dyson}, p. 635). His article started with Ramanujan’s $\tau$-function, dealt with several group theoretical topics and led to an appeal for  an integration of quantum field theory and Riemannian geometry. We can do no better than follow his guidance. Whatever else is common between Ramanujan, Galois and Riemann, in each case it took a long time until the importance of their ideas for an understanding of nature was recognized. The circumstances were very different. In the intellectual environment of France, contact with physics was unavoidable. Galois studied Jacobi’s work on elliptic integrals and he recognized deep relations with his own theory of algebraic equations. But this was an aside. His study of finite groups owed nothing to physics. Only after the discovery of quantum mechanics its importance for the understanding of nature became obvious. Nevertheless, this development was not particularly surprising. Symmetry is a very fundamental concept, in nature, art and mathematics. Crystal symmetries had been studied before Galois. Even more importantly, the relation between continuous symmetries and conserved quantities underlies Kepler’s second law and is basic for much of classical mechanics. Quantum physics uses representation theory for a common perspective on those rather diverse groups. Galois did not know it yet, but fortunately it had been developed for internal mathematical reasons when it was needed in physics. Of particular importance was the use of tensor products by Schur. When two independent physical systems in quantum mechanics are described by vector spaces $V_1, V_2$ the compound system is described by $V_1\otimes V_2$. This simple fact is at the core of our first axiom. In the near future it should become the basis of quantum computation.

In hindsight, one can make a very simple argument. The outcome of measurements is described by real numbers, so maps to real vector spaces must be important. This argument glosses over a rather mysterious aspect of quantum theory, namely the fact that nature uses a projective vector space over the complex numbers. As a result, one needs representations of central extensions of many groups. For euclidean QFT this aspect has no immediate relevance, but we will see a related extension in the next paragraph. For a start, we just use a simple rule of thumb - for any mathematically interesting set with relevance to nature, one should look at its maps to the real rumbers. We will apply it to Riemannian manifolds. Orientation and spin structures will not be considered, though they come up in deeper investigations. 

In contrast to Galois, Riemann had enough time to develop his interest in physics. In hindsight, he was looking for Maxwell’s equations, but did not succeed. Instead, he developed Riemannian geometry and hoped that it would become important for physics at sub-microscopic scales. In a sense he was right, though the best known applications of his geometry concern astronomy and cosmology. We will make much use of the basic operations he introduced, cutting and pasting along boundaries, Cartesian products, diffeomorphisms and transformations of the metric that preserve the conformal structure. At least in physics, the latter are called Weyl transformations, because Weyl wanted to use this abelian group for a theory of electromagnetism. Local transformations of the metric modulo Weyl transformations yield transformations of the conformal structure. In two dimensions, the latter is best described by coordinate functions that satisfy the Cauchy-Riemann equations. 

Here are some notations that will be used. Non-degenerate symmetric bilinear forms will be called scalar products.  Generic fields will be denoted by $\Phi$, often with indices. The usual symbol for Riemannian manifolds will be $M$, often with indices. Symbols for manifolds that have a boundary include a hat, like $\hat M$. Boundaries will be smooth or empty and denoted by $B$. More generally, we say that $\hat M$ has boundary $B$ when it is provided with an isometric map of its boundary to $B$. ${\mathcal B}^n$ and $S^{n-1}$ are the unit ball and the unit sphere in $\RR^n$. 
A hat would be overkill and calligraphic B is used, because boundary and ball begin with the same letter. The Riemannian metric on $M$ will be denoted by $\mu_M$. When it has to be written in components we use the standard notation $\mu_M =g_{\mu\nu}dx^\mu dx^\nu$, with the suppression of the summation sign introduced by Einstein. The Cartesian product $B\times [0,T]$ has the standard metric $(dt)^2+\mu _B$ for $0<t<T$. Positive real numbers $\lambda$ act on metric spaces by multiplication of the distances, as in $\lambda M$. These maps are called scale transformations. For Riemannian manifolds, $\mu_{\lambda M}=\lambda^2\mu_M$. $\lambda{\mathcal B}^n$  is the ball of radius $\lambda $ around the origin in $\RR^n$ and $\lambda S^{n-1}$ is its boundary.  For $a\in\RR^n$, $a+\lambda {\mathcal B}^n$ is the translated ball with center $a$.

For $n\in\NN$ we define a manifold category ${\mathcal M}(n)$ as follows. The objects of ${\mathcal M}(n)$ are the smooth compact Riemannian manifolds of dimension $n-1$ and the empty set. The arrows from $B_i$ to $B_o$ are compact Riemannian manifolds $\hat M$ of dimension $n$ that have boundaries with a partition $B=B_i\sqcup B_o$ into an incoming and an outgoing part. When such an $\hat M$ is smooth it is an arrow of ${\mathcal M}(n)$. Arrows $\hat M^1$, $\hat M^2$ can be composed in the obvious way whenever $B_o^1=B_i^2$. ${\mathcal M}(n)$ is the smallest category with these properties. It is monoidal with respect to the disjoint union. The isomorphism classes of those arrows of ${\mathcal M}(n)$ that have empty boundary form a metric space ${\mathcal R}(n)$. Its completion is denoted by $\bar{\mathcal R}(n)$. The curvature of a Riemannian manifold of ${\mathcal M}(n)$ has a smooth part and a part supported on a codimension 1 submanifold.

As a metric space, ${\mathcal R}(n)$ has a natural topology. Any euclidean QFT can be characterized by a continuous map $Z:{\mathcal R}(n)\rightarrow \RR$. These maps are called partition functions. $Z$ is determined by its values for smooth manifolds, but we add to its domain those points of $\bar{\mathcal R}(n)$ at which it has a unique limit. 
Continuous families of QFTs can be defined by continuous maps $\RR\times {\mathcal R}(n)\rightarrow \RR$. The study of such families is called perturbation theory. In particular, composing $Z$ with scale transformations yields the renormalization group flow in the space of QFTs. The trivial QFT in dimension $n$ is defined by $Z(M)=1$ for all $M\in{\mathcal R}(n)$. It belongs to a continuous family of QFTs with partition functions of the form $Z(M)=e^{\Lambda V(M)}$. Here $V(M)$ is the volume of $M$ and $\Lambda$ is a real number called the cosmological constant. 

Models can be multiplied and added by multiplying their partition functions, or by adding them when $M$ is connected. From a physics point of view, addition yields physics in different worlds, whereas multiplication yields several kinds of matter in the same world, without interaction between them. We always will assume that a QFT is irreducible with respect to addition. Products are far more interesting, because perturbations can introduce interactions. For topological QFTs the partition function is constant in each connected component of ${\mathcal R}(n)$. We only will need the case $n=2$, for which $Z(M)$ is obtained by exponentiating the Euler characteristic of $M$. In a scaling QFT, composing  $Z$ with a scale transformation just yields a multiplication of $Z$ with the partition function of a topological QFT. In conformal QFTs, this property has a partial generalization to arbitrary Weyl transformations. Here $Z$ is multiplied by a simple factor that only depends on the local geometry of $M$. If this factor is different from 1, it is called a conformal anomaly. The Rogers-Ramanujan QFT and the models that yield mock modular forms are two-dimensional and conformal.

We will use two axioms, namely the Segal-Atiyah axiom and conformality. The Segal-Atiyah axiom is essentially the one used in topological QFT, with the addition of smoothness. It arose in G. Segal’s attempts to find an appropriate mathematical language for conformal field theory in two dimensions. His ideas had a deep impact on the physics community, but they suffered from the attempt to avoid any reference to metrics. Before their publication \cite{Segal}, they circulated for many years as a preprint. They are more natural than the related ones of vertex operator algebra and carry over to generic QFTs, though an aspect of functional analysis still needs a better understanding. Atiyah extracted what was the mathematically most transparent part and used it for his creation of topological QFT \cite{Atiyah}. 

Most of the interesting properties of a QFT relate to a vector space of fields that is filtered in terms of the so-called scaling dimension. Intuitively, fields can be conceptionalized as the differential operators
admitted by $Z$ and their scaling dimensions as the order of the operator, though the order is not necessarily integral. One particularly important field is given by the derivative with respect to the metric. More precisely, this is a tensor of fields, with components corresponding to the components of $\mu$. We shall call it the Hilbert field, because of \cite{Hilbert}, p. 404. Hilbert made this observation in the frantic and fruitful months when he worked hard to have an impact on Einstein’s emerging theory of gravity and Einstein worked hard with the opposite aim. Hilbert’s insight concerned classical field theory, but carries over immediately to QFT. Its reception in physics is a complicated story and would need a historian to untangle. For flat manifolds the Hilbert field can be identified with the energy-momentum tensor.  For conformal QFT in two dimensions, the Hilbert field components are the real and imaginary parts of the Virasoro field, up to scaling by $2\pi$.

The local multiplicative behaviour of fields is described by the  operator product expansion (OPE), global properties are expressed by $N$-point functions. In contrast to the more complicated Minkowskian situation, the canonical fields of euclidean QFT do commute. Fields, $N$-point functions and the OPE will be defined in terms of Segal-Atiyah data, but one needs some asymptotic scaling property of $Z$ to guarantee that the definition produces the tapestry one is used to. Conformality is more than sufficient for this purpose, so that we can focus on this special case. The combinatorial property expressed by the sum form of the Rogers-Ramanujan functions follows immediately from the OPE of the Virasoro field. An explanation of the product form would require the study of non-conformal perturbations and will not be attempted.

\section {A brief introduction to euclidean quantum field theory}

For a start, one can think about QFT as the study of functions $Z:{\mathcal R}(n)\rightarrow \RR$ that share natural properties with the functions $ Z^{(n)}$ defined by
$$ Z^{(n)}(M)=(\det(\Delta_M+m^2))^{-1/2}.$$
Here $\Delta_M$ is the positive Laplace operator of $M$. It has a kernel given by the constant functions, also called zero mode. A mass $m\in\RR_+$ has been introduced to avoid a division by zero. The determinant may be defined using $\zeta$-function regularization (\cite{Ray-Singer}, \cite{Hawking}).

Except in easy cases, the author does not know, which of the properties used below have been checked for $ Z^{(n)}$. Instead, he relies on physics folklore. One property of $ Z^{(n)}$ is that its values are positive. This property is not shared by the Rogers-Ramanujan model, so we do not want to impose it. Nevertheless, in that model and in many other QFTs the range of the partition function does not include 0. This may well be true in general.

To stay close enough to Ramanujan’s ideas, we will defer the statement of the axioms and first calculate some values of $Z^{(n)}$. Let $B\in {\mathcal R}(n-1)$. We calculate $Z^{(n)}(B\times TS^1)$ like Euler, to get to the point without much technical effort. Let $E_i^2$ be an eigenvalue of $\Delta_B +m^2$, with $E_i>0$. Its contribution to the determinant is
$$\prod_{k\in\ZZ}(E_i^2 + (k/T)^2).$$ 
Regularization yields $\prod_{k\in\ZZ} T = 1$, because $\zeta(0)= -1/2$. Thus the regularized product over $k$ yields $(\sinh (\pi E_iT))^2$, up to a numerical factor. Altogether one obtains 
$$Z^{(n)}(B\times TS^1) = \exp(-2\pi E_C T) \prod_i (1-e^{-2\pi E_iT})^{-1},$$
where $2E_C$ is the regularized sum of all $E_i$. In physics terminology, $Z^{(n)}(B\times TS^1)$ can be interpreted as the trace of an operator $e^{-2\pi(H+E_C)T}$ on a space ${\mathcal S}(\oplus_i V_i)$, where the $V_i$ are eigenspaces of $H$ for eigenvalue $E_i$ and ${\mathcal S}$ denotes the symmetric tensor algebra. After continuation to Minkowskian space, the restriction to the $k$-fold symmetric product of $\oplus_i V_i$ describes the quantum mechanics of $k$ indistinguishable and non-interacting particles, with the positive squareroot of $\Delta_B +m^2$ as Hamilton operator for each particle. For $k=0$ one obtains the vacuum. The summand $E_C$ is called the energy of the vacuum, or the Casimir energy, because Casimir discovered it in 1948 in the related case of Maxwell’s equations. One often uses the notation $q=e^{-2\pi T}$ and writes $Z^{(n)}(B\times TS^1)$ as a sum over powers of $q$ with positive integral coefficients and real exponents. When one ignores the Casimir factor, the coefficient of $q^E$ is the number of partitions of $E$ as a finite sum with terms $E_i$, repetitions being allowed. $Z^{(2)}(LS^1\times TS^1)$ is invariant under an interchange of $T$ and $L$. For small $m$ this yields the modular invariance of Dedekind’s $\eta$-function. 

This yields a first overlap with Ramanujan’s work, because he was fascinated both by partitions and by the problem of finding a natural value for divergent expressions. This aim was not particularly original. Euler worked with divergent series and got reasonable results, which later could be justified without sleight of hand. Ramanujan compared his own result for divergent sums to the center of gravity of a body (\cite{Berndt1}, p. 134). He may have felt some relation with classical physics, but a natural link only occurs in QFT. There, models involve scales that are arbitrarily small and simultaneously active. On first sight, this appears to yield divergences, but everything adds up to something finite, like the Casimir energy. In mathematics, the relevant technique was used by Weierstrass, for the construction of holomorphic functions with given zeros. In general, the obvious answer is divergent. One regularizes it by a restriction to a finite number of zeros, renormalizes it by the multiplication with a suitable exponent and takes the limit. The renormalization must be chosen so that the limit exists. When the number of zeros in a finite region satisfies a polynomial bound, the renormalization just needs the exponential of a polynomial and only depends on its coefficients. Numerical studies of QFTs are very similar. One works with a finite lattice and models anything smaller with the help of a few parameters. 

The relevance of partitions to physics was revealed in 1924 in a letter of the Indian physicist Bose to Einstein, who immediately recognized its value. In contrast to Ramanujan's somewhat earlier letter, it contained only one result and had no sequel. Still, Bose was immortalized by the word boson for particles which have a partition function like the one described above. More specifically, the model introduced above is called the QFT of a free scalar boson. Bosons correspond to the trivial representation of the permutation group, whereas the signature representation yields fermions. The QFT of free fermions uses Pfaffians of Dirac operators instead of inverse determinants. Symmetric products are replaced by exterior products, so that instead of factors of the form $(1-q^E)^{-1}$, fermions yield $1+q^E$. Number theorists will recognize a possible link with $L$-functions. 

Unfortunately, Ramanujan died prematurely, a few years before Bose’s letter. He would have had no difficulty to understand his ideas and might have anticipated the Casimir term. He knew that $\prod_{n=1}^\infty (1-q^n)^{-1}$ has to be multiplied by $q^{-1/24}$ for nice transformation formulas. Because Ramanujan's regularization agrees with $\zeta$-function regularization in this case, he may have seen that the exponent can be interpreted as $(1+2+3+\ldots)/2$. Analogously, the Rogers-Ramanujan function
$$G(q)=\prod_{n=o}^\infty\frac{1}{(1-q^{5n+1})(1-q^{5n+4})}$$
requires multiplication by $q^{-1/60}$ and its partner
$$H(q)=\prod_{n=o}^\infty\frac{1}{(1-q^{5n+2})(1-q^{5n+3})}$$
multiplication by $q^{11/60}$. These Casimir factor should come from $1+4+6+9+11+14+\ldots=-1/30$ and $2+3+7+8+12+13+\ldots=11/30$. Indeed, $\zeta$-function regularization yields the Hurwitz value $\sum_{n=0}^\infty (pn+q) = p\zeta(-1,q/p)$, thus
$$\sum_{n=0}^\infty (pn+q) = \frac{q(p-q)}{2p} -\frac{p}{12}.$$
Instead, Ramanujan’s method supposes that $\sum_{n=0}^\infty (pn+q) = p\sum_{n=0}^\infty n + q\sum_{n=0}^\infty 1,$ so that it misses the term $-\frac{q^2}{2p}$. Just possibly, Bose’s letter might have put him on the right track. Or is this just the dream of a physicist?

In physics, the basic idea of QFT is that local structures of limited complexity integrate up to big things like protons, water or humans. In practice, this idea may be expressed by writing $Z$ as a functional integral. Though such integrals do not have an established mathematical definition, there is a large and important community of physicists that approximate some of them by lattice calculations of ordinary integrals. The basic procedure is simple and serves well as motivation for the Segal-Atiyah axiom. In the case of $Z^{(n)}$ one writes
$$Z(M)=\int \exp\left(-\int_M ((\nabla\phi)^2+m^2\phi^2) dV_M\right) d\phi.$$
Here $dV_M$ is the volume form of $M$. Formally, $\phi$ varies over maps from $M$ fo $\RR$. The formal expression can be regularized by approximating $M$ by a finite number of points and the derivatives by differences for points at close distance. Any such difference yields a link between points, so that $M$ is approximated by a graph $M^f$. The function $\phi$ restricts to a function $\phi^f$ on the chosen points, so that the integral becomes finite dimensional and yields some number $Z(M^f)$. One would like a choice of the normalization of the measure $d\phi^f$ so that $Z(M^f)$ converges to $Z(M)$ when $M^f$ is refined. The choice should depend on local data only and keep track of the scalar product
$$(\phi_1,\phi_2)=\int_M \phi_1\phi_2\ dV_M.$$
Most likely, this would have been an easy question for some of the great mathematicians of the past.  Quite possibly, the answer exists in the mathematical literature, but has not spread to the lattice community. It would be worth many millions. In daily practice, $M$ is a rectangular torus with ratios of edge lengths that are rational numbers. It is approximated by rectangular grids, for which the correct normalization is easy to find. 

We shall bypass the problem by the following observations. Let $B$ be a compact codimension one submanifold of $M$. One can choose discretizations $(M-B)^f$ and $B^f$ so that no link crosses $B$. Now $M-B$ is a manifold $\hat M$ with two copies of $B$ as boundary. When one considers the restrictions of $\phi$ to $B^f\sqcup B^f$ as variable and integrates over the restriction of $\phi$ to $(M-B)^f$ one obtains an element of $Q(B^f)\otimes Q(B^f)$, where $Q(B^f)$ is a real Hilbert space with an ${\mathcal L}^2(\RR)$ factor for each point in $B^f$. The value of $Z(M^f)$ is given by the scalar product $Q(B^f)\otimes Q(B^f)\rightarrow \RR$. When $B$ splits $M$ into two manifolds $\hat M_1, \hat M_2$ with boundary $B$, the corresponding separate integrations yield two vectors $v_1, v_2$ in $Q(B^f)$ and the vector $v_1\otimes v_2$ in $Q(B^f)\otimes Q(B^f)$.

Instead of worrying about the measure $d\phi^f$,  we demand that this scheme carries over to a smooth monoidal functor $Q$ from ${\mathcal M}(n)$ to a category ${\mathcal V}$ of real vector spaces. To first order, smoothness means that the dependence of $Z(M)$ on the metric of $M$ can be linearized for small distances. Higher orders might not be necessary, or might need some extra thought. Otherwise the required technicalities are standard, with the following exception. In the example just discussed the real vector spaces are Hilbert spaces. This is the defining property of unitary theories, but not true in the Rogers-Ramanujan case. Segal used
locally convex complete topological vector spaces, but self-dual Banach spaces may be more appropriate. An artificial example is the orthogonal sums of two Hilbert spaces, with the standard norm, but the duality map $(v_1,v_2)\mapsto (v_1,-v_2)$. More intrinsically, our construction yields the set of vectors $S(B)$ in $Q(B)$ that can be written in the form $Q(\hat M)1$, where $\hat M$ has boundary $B$. A natural candidate for a norm in $Q(B)$ is given by
$$||v||^2=max\left\{\frac{\ |(v,w)|^2}{|(w,w)|}\ \Big\vert\ w\in S(B)\right\},$$ 
but this needs further study.

For manifolds, monoidal means that ${\mathcal M}$ contains the empty set and admits disjoint unions, for vector spaces that ${\mathcal V}$ contains $\RR$ and admits tensor products. For $Q$ one needs  $Q(\emptyset)=\RR$ and $Q(\hat M^1\sqcup \hat M^2) = Q(\hat M^1)\otimes Q(\hat M^2).$ 
 
\begin{Def}
A monoidal functor $Q$ from ${\mathcal M}$ to ${\mathcal V}$ is self-adjoint, if it has the following property.
Let $B=(\tilde B_i\sqcup B^c) \sqcup \tilde B_o$ and $B= \tilde B_i\sqcup (B^c\sqcup \tilde B_o)$ be different bipartitions of the boundary $B$ of a given $\hat M$. Then the corresponding maps $Q(\tilde B_i)\otimes Q(B^c)\rightarrow Q(\tilde B_o)$ and $Q(\tilde B_i)\rightarrow Q(B^c)\otimes Q(\tilde B_o)$ are adjoint with respect to the scalar product on $Q(B^c)$. 
\end{Def}

We now can formulate the Segal-Atiyah axiom for partition functions. We do not call it a definition, because the functional analysis in the non-unitary case is incomplete.
 
\begin{Axiom} When $Z$ is the partition function of an $n$-dimensional QFT, there is a self-adjoint functor $Q$ from ${\mathcal M}(n)$ to a category ${\mathcal V}$ of self-dual Banach spaces so that $Q(M)1=Z(M)$ for $M\in {\mathcal R}(n)$. 
\end{Axiom}

We state some crucial consequences. For $B\times [0,T]$ we can use the identity map for the incoming and any isometry of $B$ for the outgoing boundary. In the limit $T\rightarrow 0$ this yields an action of the isometry group of $B$ on $Q(B)$.
The multiplication map $Q(B)\otimes Q(B)\rightarrow\RR$ is an adjoint of the identity and describes the glueing of two outgoing boundary components that are both parametrized by $B$. The dual adjoint yields a comultiplication, which describes the glueing of incoming boundary components. The arrows $B\times [0,2\pi T]$ yield a semigroup of symmetric operators in the endomorphism semigroup of $Q(B)$, with trace $Z(B\times TS^1)$. Thus one can write 
$$Q(B\times [0,T])=\sum_I\exp(-2\pi E_I(B)T)P_I,$$ 
where the $P_I$ are orthogonal projectors to finite dimensional subspaces of $Q(B)$. This ties in with the calculation of $Z^{(n)}(B\times TS^1)$ given above. The letter $Q$ expresses the property that one obtains a quantum mechanical system on $Q(B)$.  with eigenvalues $E_I(B)$ of its Hamiltonian for $B$. For $n=1$, partition functions are cosmologically constant or additively reducible.

Because $Q$ is monoidal, $Z(M_1\sqcup M_2) = Z(M_1)Z(M_2)$. One can ask, if this formula remains true when $M_1$ and $M_2$ intersect in one point. Conjecturally, a positive answer is equivalent to unitarity. In any case, for unitary QFTs the range of $Z$ is positive.

Consider the family ${\mathcal M}(B)$ of manifolds $\hat M$ for which the incoming boundary is empty and the outgoing boundary $B$ is fixed. The span of the spaces $Q(\hat M)1$ yields a subspace $Q_b(B)$ of $Q(B)$. When $B$ has the topology of a sphere, a subspace $Q_v(B)$ of $Q_b(B)$ is obtained from those manifolds in ${\mathcal M}(B)$ that have the topology of a ball. One calls $Q_v(B)$ the vacuum sector of $Q(B)$. For the Rogers-Ramanujan model we shall see that $Q_b(B)=Q(B)$, whereas $Q_v(B)$ is a proper subspace. To gain some insight into the mock modular phenomenon, we need models for which $Q_b(B)$ is a proper subspace of $Q(B)$.

The vector spaces $Q_b(B)$ are determined by $Z$ up to unique isomorphy, as follows. Let $\hat M_1$ and the dual of $ \hat M_2$ lie in ${\mathcal M}(B)$. For $M(T)=\hat M_1\cup B\times [0,T] \cup \hat M_2$ one can write
$$Z(M(T)) = \sum_I A_I \exp(-2\pi E_I(B) T).$$
For each $I$, the coefficient $A_I$ yields a scalar product on a vector space spanned by one basis vector $\vert \hat M\rangle$ for each  $\hat M\in{\mathcal M}(B)$. Taking the quotient with respect to the subspace of null vectors yields a finite dimensional vector space that can be identified with  $P_IQ_b(B)$. For analogous ideas in topological QFT and their history, see \cite{Khovanov}. In physics terminology,  $Q_b(B)$ is the basic superselection sector of $Q(B)$. The dimension $D_I$ of $P_IQ(B)$ can be read off from $Z(B\times TS^1)$,
so that it is easy to check if $Q_b(B)=Q(B)$. 

If $Q(B)$ contains additional vectors, one can proceed as follows. Let $B$ be cobordant to $B’$. In general, $Q_b(B\sqcup B’)$ is larger than $Q_b(B)\otimes Q_b(B’)$. In particular, for each vector $v$ in $Q(B)$ the vector space $Q_b(B\sqcup B)$ includes the tensor product of $v$ with a proper subspace of $Q(B)$. Nevertheless, this does not mean that $v$ itself can be determined up to unique isomorphy. Let us show that for $Z^{(n)}$. 
The involution $\sigma$ given by $\phi\mapsto -\phi$ acts on $Q(B^f)$ and in the limit on $Q(B)$. This yields a splitting $Q(B)=Q_+(B)\oplus Q_-(B)$. The discretized Laplace operator commutes with $\sigma$, so that $Q_b(B)$ is contained in $Q_+(B)$. It turns out that $Q_b(B)=Q_+(B)$. Let $B=B^1\sqcup B^2$. Then
$$Q_b(B)=(Q_b(B^1)\otimes Q_b(B^2))\oplus (Q_-(B^1)\otimes Q_-(B^2)).$$
For $v\in Q_-(B)$ there is a canonical construction of $v\otimes v$, but no canonical way to distinguish $v$ from $-v$. Isomorphisms like $\sigma$ yield the superselection group of any QFT. They can be read off from $Q$ by Tannaka-Krein duality \cite{DoplicherRoberts}. It is natural to generalize $Z(B\times TS^1)$ to functions
$$Z_B(g,B\times TS^1)=\tr (gQ(B\times [0,2\pi T])),$$
where $g$ represents the superselection group of the QFT. As an example,
$$Z^{(n)}_B(\sigma,B\times TS^1) = \exp(-2\pi E_C T) \prod_i (1+e^{-2\pi E_iT})^{-1}.$$

With this detour we have come closer to Ramanujan’s ideas again. Partition functions yield series with positive integral coefficients, but the inclusion of superselection group elements leads to arbitrary integral coefficients. Ramanujan considered such modifications, but he did not relate them to group theory.

The numbers $D_I$ allow to define the complexity of a theory. For definiteness, take $B=S^{n-1}$, though any other choice should give the same result. Let $I(E)$ be the set of indices for which $E_I<E$ and put $D(E)=\sum_{I\in I(E)} D_I$. For $Z^{(n)}$ this function will be called $D^n$. One finds $log\ D^n(E) = \kappa E^{(n-1)/n}$ up to terms of lower order, with some numerical factor $\kappa$. 
We define the complexity $c_{eff}$ of a QFT in $n$ dimensions as the upper limit of $(log\ D(E))/(log\ D^n(E))$ for large $E$. The notation comes from conformal QFT in two dimensions, where 
$c_{eff}$ is called the effective central charge. 

Up to scale transformations, there is most probably a unique non-trivial theory of smallest possible complexity. For this QFT, $c_{eff}=2/5$ and the restriction of $Z$ to flat tori yields the Rogers-Ramanujan functions.  Among the non-trivial conformal QFTs in two dimensions, it is not hard to prove that this model has the smallest possible complexity. It is even less complex than the QFT of a real massless free fermion, for which $c_{eff}=1/2$.

Both for physical and for mathematical purposes, QFTs should have partition functions with manageable moduli spaces with respect to perturbations. Almost certainly, this requires something more than the Segal-Atiyah axiom. Just possibly, demanding finite complexity might be sufficient, but at present this question is far too hard. Instead, note that the discretization of $Z^{(n)}$ uses graphs
with short links only. This property is called locality and has a good physics pedigree ({\it pace} Newton).
The Segal-Atiyah axiom is just about localization to submanifolds of codimension 1. One might increase codimensions in steps of 1, but in practice one often looks at points. The corresponding limit vectors are called fields.

\begin{Def}
Consider the directed system given by $\epsilon{\mathcal B}^n- \epsilon'{\mathcal B}^n$ for $\epsilon’<\epsilon$ and the corresponding maps $Q(\epsilon' S^{n-1})\rightarrow Q(\epsilon S^{n-1})$. Let $Q_0$ be its projective limit and $\pi_\epsilon$ be the corresponding projections. Let 
$$F_{\leq d}=\{\Phi\in Q_0\mid\  ||\pi_\epsilon\Phi|| = {\mathit O}(\epsilon^{-d})\}.$$
For $d\in\RR$, the elements of $F_{\leq d}$ are called fields of scaling dimension $\leq d$.
The union $F=\bigcup_{d\in\RR}F_{\leq d}$ is the vector space $F$ of fields of the QFT.
\end{Def}

The superselection group acts on $F$ and yields a canonically constructable invariant subspace $F_b$, which contains a vacuum sector $F_v$. Elements of $F^{\otimes N}$ are called $N$-local fields.
The QFT axioms should imply that $\pi_\epsilon F$ is dense in $Q(\epsilon S^{n-1})$ for any $\epsilon$. Fields lie in finite dimensional representations of the $\RR^n$ rotation group $O(n)$ 
and can be differentiated according to
$$a^\mu\pi_\lambda(\partial_\mu\Phi)+\pi_\lambda\Phi=Q(\lambda{\mathcal B}^n-(a+\epsilon{\mathcal B}^n))\pi_\epsilon\Phi+o(|a|),$$
for $\lambda>\epsilon$ and sufficiently small $a$ in $\RR^n$. For $\Phi\in F_{\leq d}$ one obtains
$\partial_\mu\Phi \in F_{\leq (d+1)}$ because of the $a^\mu$ factor on the left of this equation.
The field defined by $\pi_\epsilon\Phi=Q(\epsilon  S^{n-1})1$ is denoted by 1. It lies in $F_v\cap F_{\leq 0}$, transforms trivially under $O(n)$ and its derivatives vanish. In unitary theories, $F_{\leq d}$ is 0 for $d<0$ and spanned by 1 for $d=0$.

When a manifold $\hat M$ has metric $\mu$, let $\hat M(\mu+h)$ be the corresponding manifold with metric $\mu+h$, where $h$ must have support in the interior of $\hat M$. The Hilbert field tensor $T^{\mu\nu}$ is defined by
$$\pi_\epsilon T^{\mu\nu}=2 \frac{\delta }{\delta h_{\mu\nu}}\Big|_0 Q(\epsilon{\mathcal B}^n(\mu+h))1.$$
The notation $\vert_0 $ means that the derivative is taken at the origin and for $h=0$. Recall that the functional derivative is defined by
$$Z(M(\mu+h))-Z(M)=\int \frac{\delta Z(M)}{\delta h_{\mu\nu}}h^{\mu\nu}\ dV_M+o(h).$$
To approximate $T^{\mu\nu}$ one needs to divide by a small volume, which yields $T^{\mu\nu}\in F_{\leq n}$.

The insertion of $N$-local fields in $Z$ yields the $N$-point functions of a model, perhaps the most important tool in QFT. Let us consider the case when local coordinates $x=(x_1,\ldots,x_N)$ yield isometries from non-intersecting open sets $U_1,\ldots,U_N$ in $M$ to open sets in $\RR^n$. For $p_i\in U_i$ let $x_{ip}=x_i-x_i(p_i)$. Let $\epsilon_i {\mathcal B}^n\subset x_{ip}(U_i)$ for $i=1,\ldots,N$. Then the corresponding $N$-point function $U_1\times\ldots U_N\times F^N\rightarrow\RR$ is defined by $Q\left(M-\sqcup_i x_{ip}^{-1}(\epsilon_i {\mathcal B}^n)\right)\pi_\epsilon F^N$ for sufficiently small $\epsilon=(\epsilon_1\ldots,\epsilon_N)$. For $\Phi_1\otimes\ldots\otimes\Phi_N\in F^N$ the standard physics notation for the values of the $N$-point function is
$\langle \Phi_1(x_1)\ldots\Phi_N(x_N)\rangle_M$, where $M$ is omitted when it is given by the context. One may include $Z$ itself as a 0-point function. The insertion of the field 1 acts as the identity map. One has $\langle \partial_\mu\Phi(x)\ldots\rangle_M=\partial_\mu\langle \Phi(x)\ldots\rangle_M.$

When the local coordinates do not yield isometric maps to open sets in $\RR^n$, the definition of
$N$-point functions is harder. One has to make a local flattening by a small change of the metric, insert the fields and take a limit. In general, this may require a renormalization of the fields. The only easy case is the one of holomorphic coordinates for conformal QFTs in two dimensions. Here the flattening can be performed by an arbitrarily small Weyl transformation and the calculation of the limit is trivial. This is the only case we shall need.

Insertion of fields $\Phi_1,\Phi_2$ at points $x_1,x_2$ of $\epsilon{\mathcal B}^n$ yields an element of $Q(\epsilon S^{n-1})$. Thus one obtains the germ of a function with values in ${\mathcal L}(F\otimes F, F)$, with projections to the space of linear maps ${\mathcal L}(F\otimes F, F_{\leq k})$ for any $k$. This germ is called the operator product expansion (OPE) of the QFT.
In physics it is notated by an expansion of $\Phi_1(x_1)\Phi_2(x_2)$ as a sum of fields in some $F_{\leq k}$, with functions as coefficients. In all known cases the germ is analytic, so that the coefficient functions are uniquely determined.
This expansion describes how $N$-point functions behave when two of the $N$ points approach each other. 
For ease of notation, one often puts $x_2=0$ and uses $\Phi_2(0)=\Phi_2$. This loses no information, because the OPE commutes with translations. When $x_2=0$, then $\Phi_1=1$ yields just the identity map, whereas $\Phi_2=1$ yields the Taylor expansion of $\Phi_1(x)$. When  $\Phi_1,\Phi_2$ both lie in $F_b$, then their OPE only yields fields in $F_b$ and analogously for $F_v$. The Hilbert field lies in $F_v$ and should generate all of $F_v$ by repeated use of the OPE.

\section {Conformal QFT in 2 dimensions and the Rogers-Ramanujan functions}

It often happens that $Z(\epsilon M)$ has a limit for small $\epsilon$, in general after multiplicative renormalization by a topological factor. The limiting QFT has simple scaling properties and allows to recover $Z$ by perturbation theory. 
In a scaling QFT there are natural isomorphisms between the vector spaces $Q(\epsilon S^{n-1})$ for any $\epsilon$, so that fields can be regarded as elements of $Q(S^{n-1})$. In two dimensions one has
$$Z(\lambda M)=\lambda^{c(1-g)/6}Z(M),$$
for $\lambda\in\RR_+$ and connected $M$ of genus $g$. The real number $c$ is called the central charge of the QFT. 

Recall that $Q(B)$ can be constructed by using a vector space with elements $\vert\hat M\rangle$. The map induced by $\vert\hat M\rangle\mapsto \lambda^{c(2g-1)/12} \vert\lambda\hat M\rangle$ yields an isomorphic map from $Q(B)$ to $Q(\lambda B)$. When one composes the annulus maps $Q({\mathcal B}^2-\epsilon{\mathcal B}^2): Q(S^1)\rightarrow Q(\epsilon S^1)$ with such isomorphisms, one obtains a semigroup of trace class operators in the semigroup of endomorphisms of $Q(S^1)$. They can be decomposed into finite dimensional Jordan blocks, with eigenvalues $\epsilon^d$ of $Q({\mathcal B}^2-\epsilon{\mathcal B}^2)$. In particular, $F$ is dense in $Q(S^1)$. 

Non-trivial Jordan blocks yield logarithms, which would lead us away from our topic. To get diagonalizable maps $Q({\mathcal B}^2-\epsilon{\mathcal B}^2)$ one has to exchange the two boundary components of ${\mathcal B}^2-\epsilon{\mathcal B}^2$. When the theory is conformal, this is possible, because ${\mathcal B}^2-\epsilon{\mathcal B}^2$ can be mapped to $S^1\times [0,-\log\epsilon]$ by a Weyl transformation. This yields $E_I(S^1)=d_I-\frac{c}{12}$ and a field space $F$ canonically isomorphic to $PQ(S^1)$. This isomorphism is called the field-state identity. The term $-\frac{c}{12}$ can be interpreted as a Casimir energy (\cite{Cardy}) and one can write $F_{\leq d}=\oplus_{d’\leq d} F_{d’}$, with $F_d=0$ for negative $d$ of sufficiently large absolute value.

To construct interesting models, we start with a simple conformal QFT of complexity 1. The model $Z^{(2)}$ is close to conformality, because $\int_M (\nabla\phi)^2 dV_M$ only depends on the conformal structure. We just have to get rid of the $m^2$ term, which spoils conformal invariance. For the corresponding functional integral, the divergence for $m=0$ is due to the fact that the constant function 1 is a zero mode of $\Delta_M$ and its discretizations. Thus shifts of $\phi$ by an additive constant do not change the integrand and yield a divergent integral over the range of $\phi$. To remove this divergence without spoiling conformality, we let $\phi$ vary over maps from $M$ to $RS^1$. Instead of a divergence, we now have a parameter $R$ and topologically non-trivial solutions of $\Delta_M\phi =0$. By convention, they are called instantons, though there is nothing instantaneous about them. Subtracting an instanton if necessary and modulo the zero mode, any $\phi$ can be lifted to a periodic map from $M$ to $\RR$. Accordingly, 
$$Z(M)=\sqrt{V_M} R\ \Theta^R(M)(\det{’}\Delta_M)^{-1/2},$$
where the prime means that the zero eigenvalue is omitted. The zero mode factor is $\sqrt{V_M} R$, because the normalized zero mode is $1/\sqrt{V_M}$. The factor $\Theta^R(M)$ is the sum of
exponentials of $\int_M (\nabla\phi)^2 dV_M$ over the instanton maps. Let us evaluate $Z(M)$ for the case where $M$ is a torus in the complex plane with coordinate $z$,  fundamental periods $\lambda, \lambda\tau$, $\Delta_M = -\partial_z\partial_{\bar z}$, $dV_M=idz\wedge d\bar z$. After Poisson summation, 
$$\Theta^R=\sqrt{\Im \tau}R^{-1}\Theta_R(\tau),$$
where
$$\Theta_R(\tau) = \sum_{k,l\in\ZZ}q^{\frac{1}{4}(kR+2\pi lR^{-1})^2}\bar q^{\frac{1}{4}(kR-2\pi lR^{-1})^2},$$
with $q=\exp(2\pi i\tau)$ as usual. The determinant yields Dedekind $\eta$-functions. Altogether,
$$Z_R(\tau) = \Theta_R(\tau)/|\eta(\tau)|^2.$$
Note that the volume factor and the factor of $T$ have dropped out. More surprisingly, $\Theta_R$ does not change when $R$ is replaced by its dual $2\pi/R$. 
Because $\int_M (\nabla\phi)^2 dV_M$ is invariant under sign change and translation of $\phi$, the
superselection group of $Z_R$ contains a reflection and the rotations of $S^1(R)\times S^1(2\pi/R)$. 
Terms coming from the different representations of the rotation group are labeled by $k,l$ in the formula for $\Theta_R$.  Introducing the corresponding characters generalizes $\Theta_R$ to a real analytic Jacobi form. 

There are non-conformal perturbations of $Z_R$ that re-introduce the mass $m$. For $m\neq 0$ one can take the limit of infinite $R$ to recover $Z^{(2)}$. For $Z_R$, as for any conformal QFT,
the spaces $F_{\leq d}$ have finite dimension for any $d$. This remains true for its massive perturbations, but not in the limiting case of $Z^{(2)}$. For $Z^{(n)}$ with $n>2$ one finds that apart from the constant field 1 the field $\phi$ of lowest scaling dimension satisfies the equation $(\Delta+m^2)\phi =0$ and has scaling dimension $(n-2)/2$. It changes sign under the involution $\sigma$ so that it does not belong to $F_b$.
The component of the OPE that takes values in ${\mathcal L}(\phi\otimes\phi, F_0)$ is given by Green’s function for $(\Delta+m^2)$. For $n=2$, Green’s function has a logarithmic behaviour that is responsible for the somewhat exceptional properties of $Z^{(2)}$.

One can construct a variant of $Z_R$ by using maps $\phi$ to $S^1(R)$ modulo point inversion, in other words maps to an interval of length $R/2$ with suitable bounday condition. In this context, $\Theta_R$ first appeared in the physics literature (\cite{Zam2}, 1990, eq. (8.170)). If one applies this construction to the Leech torus point of the moduli space of perturbations of $Z_R^{24}$, its superselection group yields the FG monster, perhaps the most magnificent fruit on the tree planted by Galois. We will remain closer to Ramanujan’s path, however.

Since $\Theta^R$ only depends on the complex structure, the conformal anomaly comes from
$(\det’\Delta_M)^{-1/2}$. For general $c$ the anomaly is the one provided by
$(\det’\Delta_M)^{-c/2}$. For convenience, we use that fact for a definition.

\begin{Def}
In two dimensions, conformal QFTs are characterized by the following property. Consider two metrics $\mu_a,\mu_b$ on $M$ with $\mu_b=e^{\sigma}\mu_a$ and $\sigma:M\rightarrow\RR$. Let $K_a,K_b$ be the corresponding Gauss curvature forms. Then
$$Z(M_b)=\exp \left(\frac{c}{48\pi}\int \sigma (K_b+K_a)\right)Z(M_a).$$
\end{Def}

When $\sigma$ is constant, the conformal anomaly becomes $\exp (\frac{c}{6}\sigma)$, as used above. 
Weyl transformations are generated by $g_{\mu\nu}\frac{\delta}{\delta g_{\mu\nu}}$, which yields a multiple of the Gauss curvature. For the field $T^{\mu\nu}$ this implies $T^\mu_\mu=0$, because $K=0$ on $\RR^n$. Because $T^{\mu\nu}$ is symmetric, this means that this tensor only has two independent components. By definition, $Z(M)$ does not change for reparametrizations of $M$.
This implies $\partial_\mu T^{\mu\nu}=0$, which are just the Cauchy-Riemann equations. By complexification, one obtains a holomorphic field $T$, called the Virasoro field and its complex conjugate $\bar T$. 

In physics one often uses the notation
$$2\frac{\delta}{\delta g_{\mu\nu}}Z(M)=\langle T^{\mu\nu}\rangle_M$$
for arbitrary $\mu_M$ and arbitrary coordinates $x$ and calls $T^{\mu\nu}(x)$ the energy-momentum tensor. In our case this yields
$$T^\mu_\mu=\frac{c}{24\pi}R,$$
where $R$ is the Riemann curvature. On the other hand, the trace of the Hilbert field vanishes and its insertion in the partition function can only yield 0. Without a terminological distinction between the Hilbert field and the energy-momentum tensor this would be confusing.

The OPE of two holomorphic fieldx only yields holomorphic fields.
The scaling dimension of the Virasoro field is equal to 2. Accordingly, one finds
$$T(z)T(z’) = \frac{c/2}{(z-z’)^4} + \frac{T(z)+T(z’)}{(z-z’)^2} + R(z,z’),$$
where $R(z,z’)$ is regular. For $z’=0$ this yields
$$T(z)T = \frac{c/2}{z^4} + \frac{2T}{z^2} + \frac{\partial_zT}{z}
+\sum_{n=0}^\infty z^n  R_n,$$
where $R_n\in F_{n+4}$. The reader may calculate how $T(z)$ must transform under holomorphic coordinate changes to preserve the form of the operator product expansion. Hint: use the Schwarzian derivative. The map from cylinder to annulus is given by $\tilde z = \exp(z)$, which requires
$$T(z)=\tilde z^2 \tilde T(\tilde z)-\frac{c}{24}.$$ 
Here the result of the insertion of the Virasoro field is denoted by $T(z)$ when the coordinate $z$ is used and by $\tilde T(\tilde z)$ when $\tilde z$ is used.
The sum of the constant terms for $T$ and $\bar T$ yields the Casimir energy.

The fields $R_0$ and $\partial^2 T$ span $F_v\cap F_4$. To achieve a theory with minimal complexity these two fields have to be proportional. One easily checks that this only can happen for  $c=-22/5$ and
$$R_0=-\frac{1}{5}\partial^2 T(z).$$
These equations define the Rogers-Ramanujan QFT. To calculate the corresponding $F$ one either can use a Lie algebra that is encoded by the singular terms of the OPE, or a combinatoric argument.

For holomorphic $\Phi^h$ the Laurent coefficients of the OPE map $F\rightarrow \Phi^h(z)F$ yield a Lie algebra action on $F$.
For $L_n\Phi=\oint_C z^{m+1}T(z)\Phi dz$, Cauchy’s theorem yields the Virasoro algebra
$$[L_m,L_n] = (n-m)L_{m+n}+\frac{c}{12}(n^3-n)\delta_{m+n,0},$$
and analogously for the Fourier components $\bar L_n$ of $\bar T$. Moreover, $L_m$ and $\bar L_n$ commute for all integers $m,n$, because there is no candidate for a singular term in $T(z)\bar T(\bar z’)$. The eigenvalues of $L_0, \bar L_0$ are denoted by $h,h'$, the corresponding subspaces of $F$ by $F_{hh’}$. One finds $h+h'=d$. Because $L_0-\bar L_0$ generates the $SO(2)$ rotations of $F$, $h-h'$ must be integral for fields in $F_b$.
We have written the indices of $L$ so that $L_mF_{hh’}\subset
F_{m+h,h’}$ and $\bar L_mF_{hh’}\subset F_{h,m+h’}$. Victor Kac introduced the opposite sign for minor historical reasons, but regretted it later.

When $h,h'$ are interpreted as weights, $F$ decomposes into irreducible lowest weight representations of the Virasoro algebra. Generic lowest weight representations yield $c_{eff}\geq 1$, but Kac found that for special values of $c$ smaller values are possible \cite{Kac}. Those values are parametrized by reduced fractions $p/q$ between 0 and 1 and with $p>1$, namely
$$c=1-6\frac{(p-q)^2}{pq}$$
and
$$c_{eff}=1-\frac{6}{pq}.$$
The corresponding QFTs are called $(p,q)$ minimal models. Among them, the unitary ones are those with $q-p=1$. The case $p/q=2/3$, $c=0$ yields the trivial QFT. The case $p/q=3/4$, $c=1/2$ yields a unitary free fermion theory. The smallest value, namely $c_{eff}=2/5$, arises for $p/q=2/5$, which yields $c=-22/5$. In this case there are just two different small irreducible representations, on vector spaces $V_0$ and $V_{-1/5}$. The index denotes its the minimal weight. Let $D_0(h)$ and $D_{-1/5}(h)$ be the dimensions of the subspaces of weight $h$. For the Rogers-Ramanujan model, $F_v$ is isomorphic to the real subspace of $V_0\otimes \bar V_0$, moreover
$$\tr q^{L_0}\bar q^{\bar L_0} = |H(q)|^2,$$
with the Rogers-Ramanujan function $H$. The technicalities are rather demanding and yield $H$ as quotient of a well-known $\theta$-series and Dede\-kind’s $\eta$-function, apart from the Casimir factor.

The proportionality of $R_0$ and $\partial^2 T$ yields a much simpler combinatorial description of $V_0$, indeed the one used by Ramanujan. The Virasoro OPE implies that $V_0$ is spanned by regularized differential polynomials in $T$. The regularization is irrelevant for counting and will be neglected in the notation. The scaling dimension of $\partial^kT$ is $k+2$. For the (2,5) minimal model and for that model only $TT$ is proportional to $\partial^2 T$. Differentiation shows that products of the form $\partial^k T\partial^k T$ or $\partial^k T\partial^{k+1} T$ are linear combination of simpler ones. Thus $D_0(h)$ counts partitions of $h$ in which 1 does not occur and for which any two terms have a difference greater than 1.

Now consider a field $\Phi$ in $F_{hh’}$. When $\Phi$ is a lowest weight field, the OPE with $T$ has the form
$$T(z)\Phi = \frac{h}{z^2}\Phi + \frac{1}{z}\partial_z\Phi+{\mathit regular\ terms}.$$
For the counting we neglect regularizations, as above. Using the proportionality of $TT$ and $\partial^2T$ one finds that the leading regular term is proportional to $\partial_z^2\Phi$ and that $T\partial_z\Phi$ is proportional to $(\partial T)\Phi$. Thus $D_{-1/5}\left(h-\frac{1}{5}\right)$ counts partitions of $h$ in which 1 may occur and for which any two terms have a difference greater than 1. Altogether one obtains
$$Z(LS^1\times TS^1)=|q^{-1/60}G(q)|^2+ |q^{11/60}H(q)|^2,$$
where $q=\exp(2\pi iT/L)$ and $G,H$ are the Rogers-Ramanujan functions. The invariance of this function under interchange of $L,T$ was of course well known to Ramanujan, but now we can interpret the exponents in terms of QFT, with $11/60 = -c/24$ and $-1/60=-c_{eff}/24$. These relations can be understood as follows. All manifold $M$ of genus 0 are conformally equivalent, so that the calculation of $Z(M)$ is given by the conformal anomaly alone. For $M(T)=\hat M_1\cup S^1\times [0,T] \cup \hat M_2$ one calculates that $Z(M(T))$ is of order $e^{cT/12}$ for large $T$. This also yields the contribution of the vacuum sector to $\tr Q(S^1\times TS^1)$. On the other hand $\tr Q(S^1\times T^{-1}S^1)$ is proportional to $e^{c_{eff}T/12}$ by the standard Tauberian calculation. In unitary theories the vacuum sector yields the dominant contribution for large $T$, which implies $c=c_{eff}$. Together with conformal invariance,
this means that in the Rogers-Ramanujan QFT $Z(M)$ is well understood and easily computable for arbitrary $M$ of genus 0 or 1.

All of this generalizes to the $(2,2k+1)$ minimal models, where a $k$-fold product of $T$ can be simplified. This leads to the Andrews-Gordon generalization of the Rogers-Ramanujan magic \cite{AG} and has links with dilogarithm identities and algebraic $K$-theory \cite {N}. Ramanujan had a strong interest in the dilogarithm, but this field had been worked on for too long to allow him to find new results.

Let us conclude this section with differential equations for the Rogers-Ramanujan functions, using now both the singular and the leading regular parts of the Virasoro OPE. When $M$ is a flat torus with period ratio $\tau$ and $Z$ is conformal, the definition of the Hilbert field yields 
$$\partial_\tau Z =\frac{1}{2\pi i}\oint \langle T(z)\rangle dz,$$
$$\partial_\tau \langle T(z)\rangle =\frac{1}{2\pi i}\oint \langle T(z)T(z’)\rangle dz’.$$

Since $\langle T(z)\rangle$ is a holomorphic functions on a torus and such functions are constant, we write $\langle T(z)\rangle =  \langle T\rangle$. The operator product expansion for the (2,5) minimal model yields
$$\langle T(z)T(z')\rangle = \frac{c}{12}\left(\partial_z^2{\mathcal P}(z-z')
-\frac{2\pi^4}{15}E_4(\tau)\right)Z+2{\mathcal P}(z-z’) \langle T\rangle,$$

where we have used the Laurent expansion of ${\mathcal P}(z)$. Indeed, the difference of the two sides is a holomorphic function of $z$ on the torus that is divisible by $z-z’$, thus zero. With $c=-22/5$ the two equations can be combined to an ordinary differential equation of order 2, of a type studied in \cite{KZ}. With
$$\oint {\mathcal P}(z)dz=-\frac{\pi^2}{3}E_2$$
one finds immediately
$$\left(q\frac{d}{dq}-\frac{1}{6}E_2(\tau)\right)q\frac{d}{dq}Z= \frac{11}{3600} E_4(\tau)Z(\tau).$$
The functions $E_2$ and $E_4$ are the standard Eisenstein series and the Serre derivative
$q\frac{d}{dq}-\frac{k}{12}E_2$ transforms modular forms of weight $k$ to forms of weight $k+2$. The coefficient $\frac{11}{3600}$ is special, because the equation becomes case XI of Schwarz' list \cite{Schwarz}, when it is rewritten in terms of the variable $j(\tau)$. The two functions $q^{-1/60}G$ and $q^{11/60}H$ form a natural basis of the solution space. They yield a vector valued representation of $SL_2(\ZZ)$, the mapping class group for genus 1. 

This is very classical, but conformal QFT implies the new fact that the approach can be generalized to all Riemann curves. For hyperelliptic ones of genus $g$ it is convenient to introduce the ramification points to describe $M$. For $g=1$ this is equivalent to the use of $j(\tau)$. For higher genus the resulting vector valued representation of the mapping class group has dimension ${\mathcal F}_{2g+1}$, where the ${\mathcal F}_k$ are the Fibonacci numbers. With respect to the position of any ramification point of $M$, the corresponding functions satisfy linear ordinary differential equations of order ${\mathcal F}_{2g+1}$ with moderately simple coefficients.

\section{Mock Modular Forms and some Speculations}

Mock modular forms appear in unitary supersymmetric conformal QFTs in two dimensions, for complexity 6 or higher. Though no qualitatively new mathematics is necessary for an exposition, this complexity  would require more pages. Only one example will be sketched. A stepping stone is the product of four models of type $Z_R$, as decribed in the previous section and four models of free fermions. This yields $c=4+4/2$, thus the correct complexity. When one looks at conformally invariant supersymmetric perturbations, one finds a 16-dimensional moduli space, locally given by $SO(4,4)/((SO(4)\times SO(4))$. More interesting mathematics arises, when one considers tori modulo a point reflection. This yields models linked to Kummer $K3$ surfaces and suffices for the display of mock modularity. Their perturbation yields a moduli space of 80 dimensions, including the parameters of general $K3$ surfaces. Locally it is given by $SO(4,20)/(SO(4)\times SO(20))$ and globally by the position of 4-planes relative to a self-dual lattice in 4+20 dimensions. When $M$ is a flat torus $T_\tau$ of period ratio $\tau$, $Z(M)$ must be a very special and highly symmetric function on this 80-dimensional moduli space, but it is not known yet. 

As we have seen in the case of $Z_R$, general exponents $h,h'$ of $q,\bar q$ depend on the parameter $R$, but for some special fields they stay invariant.
An invariant part of $F$ can be isolated by considering generalized partition functions that include elements of the superselection group. For $Z_R$ an $O(2)$ reflection $\sigma$ inverts the direction of the rotations. The contributions of terms with the representation labels $(k,l)$ and $(-k,-l)$ cancel in pairs, unless $(k,l)=(0,0)$. This yields
$$Z_R(\sigma,T_\tau)=\left|\frac{1}{\prod (1+q^n)}\right|^2,$$ 
a result that is independent of $R$.

The situation in the $K3$-theory is similar, but somewhat more complex. First, there are fermions, whose sign cannot be fixed canonically. Even for circle rotations of $LS^1\times TS^1$ one may or may not get sign changes. The subspace of $F$ without sign change is called the Ramond sector. Second, within this sector one can act with the superselection group. When the group element is the product of an involution $f$ that distinguishes even and odd fermion number and a certain $SO(2)$-rotation with generator $J$ one obtains a Jacobi form that is called the elliptic genus of $K3$. Written in terms of Jacobi's theta functions it has the simple form
$$Z_{Ramond}(f \exp(zJ), T_\tau)=
8\sum_{i=2}^4\left|\left(\frac{\theta_i(z,\tau)}{\theta_i(0,\tau)}\right)^2\right|.$$
Note again that it does not depend on the parameters of the model.

Now, in supersymmetric models the Virasoro OPE is enlarged by other canonical fields, in particular fermionic ones.
The irreducible lowest weight representations of the enlarged algebra come in different sizes. The partition functions of the small ones are Appell-Lerch sums and have typical mock modular behaviour. When the Appell-Lerch contribution is subtracted from $Z_{Ramond}(f\exp(zJ), T_\tau)$, one is left with a function of the form $A(q)\theta_1(z,\tau)^2/\eta(\tau)^3$, where
$$q^{1/8}A(q)=-1+45q + 231q^2 + 770q^3 + 2277q^4 +\ldots.$$
This function is mock modular. Apart from a single minus sign, its coefficients look like the dimensions of representations of the Mathieu group $M_{24}$ \cite{EOT}. Only partial explanations are known, but enough to show that a coincidence is excluded. 

The occurence of mock modular forms in supersymmetric QFT and related models of black hole physics have increased the fame and deepened the mystery surrounding Ramanujan. He already was known among physicists, because his sum $1+2+\ldots = -1/12$ explained the critical dimension of the bosonic string \cite{BN}. For a string with $d_t$ transverse dimensions one obtains a vacuum energy $-d_t/24$. For group theoretical reasons, the lowest 1-particle state must have energy 0, which yields $1-d_t/24 =0$. Adding the two dimensions of the string and its motion, one finds the required space-time dimension 26, which previously had been found in a far more complicated way. A slightly more complex argument yields the critical dimension 10 of the superstring. The connection with Riemann's $\zeta$-function was made later. Instead, physicists had rediscovered a special case of Ramanujan's thinking, as reflected in his notebooks.
 
The encounter of physics with the Rogers-Ramanujan functions was more down-to-earth. It did not happen in QFT, but in a context that may be interpreted as a discretization of a QFT,
namely in Baxter's hard hexagon model. Here is a quick description. Consider translationally invariant triangular discretizations of a two-dimensional torus. For $k=0,1,2,\ldots$ count the number of marking $k$ vertices so that no pair of marked vertices is linked and study the corresponding generating function. Baxter's treatment of this problem made extensive use of many of Ramanujan's formulas (\cite{Baxter}, section 14). The story should be told by Andrews, because of his crucial participation, for which see the reference just given and \cite{Andrews}. There is a connection with QFT, because at a critical parameter Baxter's generating function yields the partition function of the $(4,5)$ minimal model.

As exemplified by this case, it would be wrong to imagine too general a link between Ramanujan and QFT. There is a strong link with certain systems with infinitely many degrees of freedom, but what is crucial is not their smoothness, but the possibility to describe their properties in concise formulas.
Both in QFT and in discrete models the relevant property is called integrability. In classical mechanics, it distinguishes Kepler's ellipses from the chaotic orbits of the three-body problem. The latter can be studied by perturbation theory and heavy numerics, but it is hard to imagine how classical mechanics could have evolved without the early encounter with a system that is both intricate and integrable.

QFT was not so fortunate. The immediate jump from the easy case of free particles in infinite flat space to quantum electrodynamics resulted in the messy divorce commented upon by Dyson (\cite{Dyson}, p. 635). Of course, the antipathy of mathematicians was not only due to the necessity to regularize and renormalize, but also to its somewhat obscure treatment by physicists. It certainly would have helped to have mathematicians like Euler or Ramanujan as observers and perhaps as participants when QFT was developing. They certainly would have seen the gems in the garbage and might have taken their time to extract them. It took until the end of the century before quantum electrodynamics and related QFTs started to become natural in a mathematical sense, in particular by an exposition of the intricate structures coming from the combination of renormalization and perturbation theory \cite{ConnesKreimer}. The study of integrable QFTs in two dimensions took a separate path and a full merger is still in the future. If quantum gravity will have integrable features is anybody's guess.

But all that cannot explain, how Ramanujan got in touch with parts of QFT. One can come up with a rationalistic and with a romantic model. Maybe, mathematics is like a vast dark room containing all kinds of things. Mathematicians move around and notice some of them, when they get close enough. Then, step by step, they explore the neighborhood. Ramanujan moved faster than others and dared to make longer steps. In this way, he encountered some corners of integrable QFT without realizing that they are connected.
Or maybe, there is some deep connection between the unfolding understanding of nature and mathematics. Maybe this unfolding is like a piece of music where to some extent one can feel what comes next. Perhaps Pythagoras really heard that music, in the magical properties of the pentagram and in the algorithm that later was ascribed to Euclid. This is a recurrent theme, reappearing in the Fibonacci numbers, the Rogers-Ramanujan functions and the critical value of Baxter's hard hexagon model. Be that as it may, one only can regret that Ramanujan
had to develop his art quite alone. Maybe we would see much farther if he had met some Indian Galois in his school days.

\end{document}